\documentstyle[12pt,epsf]{article}
\newcount\subno      
\newcount\subsubno   
\newcount\appno      
\newcount\appeqno    
\def\nsection#1
   {
  \vskip0pt plus.1\vsize\penalty-250
    \vskip0pt plus-.1\vsize\vskip24pt plus12pt minus6pt
    \subno=0 \subsubno=0
    \addtocounter{section}{1}
    \noindent {\bf \thesection. #1\par}
    \noindent}
\def\nsubsecnn#1
   {\vskip-\lastskip
    \vskip24pt plus12pt minus6pt
    \noindent {\bf #1\par}
    \medskip
    \noindent}
\def\subsection#1
   {\vskip-\lastskip
    \vskip24pt plus12pt minus6pt
    \bigbreak
    \global\advance\subno by 1
    \subsubno=0
    \noindent {\sl \thesection. \the\subno. #1\par}
    \nobrrak
    \medskip
    \noindent}
\def\appendix#1
   {\vskip0pt plus.1\vsize\penalty-250
    \vskip0pt plus-.1\vsize\vskip24pt plus12pt minus6pt
    \subno=0
    \global\advance\appno by 1
    \noindent {\bf Appendix  #1\par}
    \bigskip
    \noindent}
\begin{document}
\setcounter{page}{1}
\pagestyle{plain}
\setcounter{equation}{0}
%
%
%
\ \\
\begin{center}
    {\large \bf Bethe Ansatz  Solution for 
 a Defect Particle in the \\[2mm] Asymmetric Exclusion Process}
\\[15mm]
\end{center}
\begin{center}
\normalsize
B.\ Derrida${}^\dag$ and    M.\ R.\ Evans${}^\ddag$ \\[5mm]
        {\it${}^\dag$ Laboratoire de Physique Statistique\\
        Ecole Normale Sup\'erieure
\footnote{laboratoire associ\'e aux Universit\'es Paris 6
Paris 7 et au CNRS}\\ 24 rue Lhomond, 
75231 Paris Cedex 05, France}\\[5mm]
        {\it${}^\ddag$ Department of Physics and Astronomy\\
        University of Edinburgh\\ Mayfield Road, Edinburgh EH9 3JZ, U.K.}
\end{center}
\noindent {\bf Abstract:} 
The asymmetric exclusion process
on a ring in one-dimension is considered with a single defect
particle. The steady state has previously been solved by a matrix
product method. Here we use  the Bethe ansatz to solve exactly for the
long time limit behaviour of
the generating function of the distance travelled by the 
defect particle. This allows us
to recover steady state properties  known from  the matrix approach
such as the velocity,
and obtain new results such as the diffusion constant
of the defect particle. In the case where the defect particle
is a second class particle we determine the large deviation function
and show that in a certain range the distribution of the distance
travelled about the mean is Gaussian. Moreover
the variance (diffusion constant) grows as $L^{1/2}$ where
$L$ is the system size. This behaviour can be related to the superdiffusive
spreading of excess mass fluctuations
on an infinite system.
In the case where the defect particle produces a shock, our expressions for the velocity and the diffusion constant coincide with those calculated previously for an infinite system by Ferrari and Fontes.
\\[6mm]
\noindent \today
\\[5mm]
\noindent Submitted to {\it J. Phys. A: Math. Gen}
\newpage
\baselineskip=18pt plus 3pt minus 2pt
\setcounter{page}{1}
\setcounter{equation}{0}
\nsection{Introduction and Model Definition}

The asymmetric simple exclusion process (ASEP) \cite{Liggett}  is a simple example of
a driven lattice gas \cite{SZ} and as such is a system far from thermal
equilibrium.  The model 
comprises particles hopping in a preferred
direction along a one dimensional lattice with hard core exclusion
imposed.  The model's broad interest lies in its connections to growth
processes, the problem of directed polymers in a random medium and
Burgers equation \cite{HHZ,KRUG}.   It is also a natural starting point for many
traffic flow models \cite{SW}.

Multi-species variants of the ASEP have been considered 
\cite{DJLS,EFGM,AHR,ADR,Karimipour}.
In particular
the idea of a second class particle has proven useful \cite{DJLS}.  The second
class particle hops forward as usual when the neighbouring site is empty
but is  overtaken by the other particles.
Therefore it moves forward in an environment of low density of particles and backwards in a high density environment.
In
this way a second class particle can be used to
locate shocks which are sudden changes in density over a
microscopic region \cite{ABL,FKS,Ferrari,FF}.
A generalisation of the second class particle idea
to that of a defect particle  \cite{BerChine,Kirone}
has been shown to exhibit phase transitions and, in particular, phase
coexistence. Interpreted in the context of traffic
problems, the phase transition corresponds to the appearance of a
traffic jam whereas coexistence between phases of different densities
corresponds to the coexistence between a freely flowing
 and a jammed region in  traffic.
 The model has also been interpreted in the context of a two way road
\cite{LPK}.

Analytical results for the ASEP, such as the diffusion constant
\cite{DEM}, and for the second class particle problem such as the
steady state current \cite{DJLS,BerChine,Kirone} have been obtained
via a matrix product technique \cite{DEHP,Krebs}.  Recently it was
shown that for the ASEP, all moments and the large deviation function
of the time integrated current can be computed via a Bethe
ansatz technique \cite{DL}.  Here we show that Bethe ansatz can also
be used in the case of the defect particle and this allows us to
generalize previous \cite{DECam,BerChine,Kirone} results obtained by
the matrix approach.

Let us now define the model we consider 
\cite{BerChine,Kirone}.
The model comprises a single defect 
particle (indicated by 2) and $M-1$ first class ({\it i.e.} normal) particles
(indicated by 1) on a ring of size $L$ sites.
The hopping rates of the particles are as follows
\begin{eqnarray}
1\ 0 &\to& 0\ 1 \;\;\;\mbox{with rate}\;\;\; 1 \nonumber \\
2\ 0 &\to& 0\ 2 \;\;\;\mbox{with rate}\;\;\; \alpha \nonumber \\
1\ 2 &\to& 2\ 1 \;\;\;\mbox{with rate}\;\;\; \beta \;.
\label{dynamics}
\end{eqnarray}
By this it is implied, for example, that
in an infinitessimal time interval $dt$ a first class
particle hops to the neighbouring site to the right with probability
$dt$ if that neighbouring site is empty.  We restrict ourselves
to $\alpha, \beta >0$.

\nsection{Main Results}
Before discussing the technical details of the Bethe ansatz solution
we summarise in this section our main results.  Let us denote by $y_t$
the distance travelled (total number of hops forward minus total
number of hops backward) by the defect particle.  In the steady state,
$y_t$ is a random variable.  Its first and second moments give the
velocity $v$ and the diffusion constant $\Delta$ of the defect
particle
\begin{eqnarray}
v &=& \lim_{t\to \infty} \frac{ \langle y_t \rangle}{t}
\label{vdef}\\
\Delta &=&
\lim_{t\to \infty}
\frac{ \langle y_t^2 \rangle-\langle y_t \rangle^2}{t}
\label{Deltadef}
\end{eqnarray}
All cumulants of $y_t$ can be computed from the knowledge of the
generating function $\langle e^{\gamma y_t} \rangle$ via
\begin{equation}
\langle y_t^n \rangle_c = \left.  \frac{d^n \ln \left[ \langle
e^{\gamma y_t} \rangle \right] } {d \gamma^n} \right|_{\gamma=0}\; .
\end{equation}

Here, by employing a Bethe ansatz technique we calculate exactly the
large $t$ behaviour of this generating function namely
\begin{equation}
\lambda(\gamma) = \lim_{t \to \infty} { \ln \left[\langle 
e^{\gamma y_t} \rangle\right] \over t}
\label{lambdadef}
\end{equation}
 for arbitrary  $L$ and $M$.
Exact expressions of $v$ and $\Delta$ follow easily from the knowledge
of $\lambda(\gamma)$
\begin{equation}
 v =  \left. {d \over d \gamma} \lambda(\gamma) \right|_{\gamma=0}
 \quad  \quad
 \Delta =  \left. {d^2 \over d \gamma^2} \lambda(\gamma) \right|_{\gamma=0}\;.
\label{Deltadef1}
\end{equation}
In the thermodynamic limit ($L,M$ large) with fixed
density $\rho$ where
\begin{equation}
\rho = M/L
\end{equation}
our exact expressions for $\lambda(\gamma)$ allow us to show  that
the velocity $v$ and diffusion constant $\Delta$ of the defect particle
have the following
asymptotic  forms in different regions of the parameter space of
$\alpha,\beta,\rho$
\begin{eqnarray}
\mbox{ For $\beta > \rho > 1-\alpha$ } & \ \ \ 
\mbox{$ v=1-2\rho$  \ \ and }&\Delta \simeq \frac{( L \pi \rho(1-\rho))^{1/2} }{4}
\label{res5} \\
\mbox{ For $\beta < \rho$  and $1-\alpha < \rho$} & \ \ \ 
\mbox{$v= 1-\beta -\rho$ \ \  and}&  \Delta =\frac{\beta(1-\beta)}{\rho-\beta}
\label{res2} \\
\mbox{ For $\beta > \rho$ and $1-\alpha > \rho$}&  \ \ \  
\mbox{$v= \alpha -\rho$ \ \  and}& \Delta =\frac{\alpha(1-\alpha)}{1-\rho-\alpha}
\label{res1} \\
\mbox{ For $\beta < \rho < 1-\alpha$} & \ \ \ 
\mbox{$v= \alpha-\beta$ \ \  and}& \Delta =\frac{\beta(1-\beta)
+\alpha(1-\alpha)}{1-\alpha-\beta}
\label{res3}
\end{eqnarray}

These results lead to a phase diagram which is essentially the same as
the one given in \cite{Kirone} with the same expressions for the
velocity.  For $\beta > \rho > 1-\alpha$ it is known
\cite{DJLS,Kirone} that the density profile, as seen from the defect
particle, has a power law decay towards its asymptotic value. In the
phases where $\beta, 1-\alpha > \rho$ and $\beta, 1-\alpha < \rho$ the
density profile decays exponentially towards the asymptotic values. In
the final phase there is coexistence between a region of low density
$\beta$ in front of the defect particle and high density $1-\alpha$
behind the defect particle. Therefore there is a shock separating the
two regions at a distance $xL$ in front of the defect particle 
where $x$ is given by $\rho = \beta x + (1-\alpha)x$
\cite{BerChine,Kirone}.  The novelty of the present work is that, as
we can calculate $\lambda(\gamma)$, all the higher cumulants of the
distance travelled (including the diffusion constant) can be
calculated exactly for the different phases.

One should notice from (\ref{res5}) that in the whole phase $\beta >
\rho$ and $1-\alpha < \rho$ (which includes the case of a second class
particle $\beta=\alpha=1$) the diffusion constant of the defect
particle increases with $L$.  This is in contrast with the diffusion
constant of the first class particles \cite{DEM,DL} which (in the
absence of a defect) decreases as $L^{-1/2}$. This difference between
first and second class particles is not a surprise since, for an
infinite system, one expects the fluctuations of position to be
superdiffusive for second class particles and subdiffusive for first
class particles \cite{vB,Spohn2}.

 Furthermore, in the whole phase  $1-\alpha < \rho < \beta$ we
will show that $\lambda(\gamma)$ defined by (\ref{lambdadef}) is given
in the large $L$ limit by
\begin{equation}
\lambda(\gamma) - \gamma (1- 2 \rho) \simeq \frac{2 \gamma }{L}
\left[ 2-\rho \left(1 + \frac{1-\beta}{\rho-\beta}
 + \frac{\alpha}{\alpha-1+\rho} \right)
\right]
 + \gamma^2 L^{1/2} {\sqrt{\pi \rho (1- \rho)} \over 8} 
\label{res6}
\end{equation}
on the scale where  $\gamma \sim O(L^{-3/2})$.

The fact that in a certain range $\gamma$, the expression 
of $\lambda(\gamma)$ is quadratic implies that
the distribution of 
 the variable 
$y_t/t - (1- 2 \rho)$ is Gaussian over a certain range. It is easy to check that  the range which corresponds to $\gamma \sim O(L^{-3/2})$ is
$y_t/t - (1- 2 \rho) \sim O(1/L)$, so that over that range,  the distribution of this difference
should be  Gaussian.

In the following sections we give the derivation of  (\ref{res5}--\ref{res6}).
\nsection{Generating function for fluctuations in distance travelled}
To calculate $\lambda( \gamma )$, we  follow and extend  the technique of references \cite{DL,DA}.
First consider $P_{\rm t}({\cal C}, y)$
which is the probability of the system 
being in configuration ${\cal C}$ and of
the defect particle having been displaced a distance $y$
(negative if the particle has been displaced backwards).
The master equation is
\begin{eqnarray}
\lefteqn{ \frac{ d\,P_{\rm t}({\cal C}, y) }{d\,t} =}\nonumber\\
&& \sum_{{\cal C}'}\left[
 {\cal M}_0( {\cal C\, , C'}) P_{\rm t}({\cal C'}, y)
 +{\cal M}_1( {\cal C\, , C'}) P_{\rm t}({\cal C'}, y-1) 
 +{\cal M}_{-1}( {\cal C\, , C'}) P_{\rm t}({\cal C'}, y+1)  \right]
\end{eqnarray}
where ${\cal M}_0( {\cal C\, , C'}),\ {\cal M}_1( {\cal C\, , C'}),\ {\cal M}_{-1}( {\cal
C\, , C'}) $ are rates for transitions from  ${\cal C'}$ to ${\cal C}$
that  do not move the defect particle, move the defect particle
forward, move the defect particle backward respectively. The total rate out of configuration ${\cal C}$ is given by  $-{\cal M}_0 ( {\cal C, C})
= \sum_{\cal C' \neq C} [ {\cal M}_1 ( {\cal C', C}) + 
 {\cal M}_0 ( {\cal C', C}) + {\cal M}_{-1} ( {\cal C', C}) ] $.
The generating function
\begin{equation}
F_{\rm t}( {\cal C}) = \sum_{y=-\infty}^{+\infty} \exp(\gamma y)
P_{\rm t}({\cal C}, y) 
\label{Fdef}
\end{equation}
obeys
\begin{equation}
 \frac{ d\,F_{\rm t}({\cal C}) }{d\,t} = \sum_{{\cal C}'}\left[ {\cal
 M}_0( {\cal C\, , C'}) F_{\rm t}({\cal C'}) + {\rm e}^{\gamma} {\cal
 M}_1( {\cal C\, , C'}) F_{\rm t}({\cal C'}) + {\rm e}^{-\gamma} {\cal
 M}_{-1}( {\cal C\, , C'}) F_{\rm t}({\cal C'}) \right]\;.
\label{Fgen}
\end{equation}
Now
\begin{equation}
\langle e^{\gamma y_t} \rangle = \sum_{\cal C} F_t( {\cal C})
\end{equation}
so we expect as in (\ref{lambdadef})
\begin{equation}
\langle e^{\gamma y_t} \rangle \sim e^{\lambda(\gamma) t}
\end{equation}
where $\lambda(\gamma)$ is the largest eigenvalue  of
the matrix ${\cal M}(\gamma) = {\cal M}_0 + e^\gamma {\cal M}_1 + e^{-\gamma} {\cal M}_{-1}$.
For $\gamma = 0$ we know that $\lambda(\gamma)= 0$
because ${\cal M}(0)$ is  a stochastic matrix. Now  as
in \cite{DL,DA}, by the Perron-Frobenius theorem
we know that the largest eigenvalue of ${\cal M}(\gamma)$ is non-degenerate
therefore as $\gamma$ increases from zero there can be no crossing
of the largest eigenvalue. Thus $\lambda(\gamma)$
is the eigenvalue that tends to zero as $\gamma$ tends to zero.

One should note in the present problem
that for $\gamma =0$ the eigenvector with eigenvalue zero
(the steady state) is non-trivial and has previously been
constructed by using a matrix product
\cite{DJLS,Kirone}. In the following sections we show how
the Bethe ansatz can recover some properties of  this steady state in the limit
$\gamma \to 0$.

\nsection{Bethe Ansatz}
\label{sec:Bethe}
Let a configuration of the particles on the ring be specified by the
co-ordinates $\{ x_1, x_2, \ldots, x_M \}$ where $x_1$ is the position
of the defect particle and $ x_2 < x_3 \ldots < x_M $ are the
positions of the normal particles. By convention, one can always
choose $\{ x_1 \ldots x_M \}$ such that $1 \leq x_1 \leq L$ and $x_1 <
x_2 < \ldots < x_M < x_1 + L $.  With the dynamics (\ref{dynamics})
the equation for an eigenfunction $\psi(x_1,\ldots x_M)$ of ${\cal
M}(\gamma)$ with eigenvalue $\lambda(\gamma)$ takes the form: for $x_i
< x_{i +1} -1 $ and $x_M < L + x_1 -1$ (i.e. when all particles are
more than one lattice spacing apart)
\begin{equation}
\lambda \psi(x_1,\ldots x_M)
=-(M-1+\alpha)\psi(x_1,\ldots x_M)
+ {\rm e}^\gamma \alpha \psi(x_1-1,\ldots x_M)
+\sum_{i=2}^{M}\psi(x_1,\ldots , x_i -1,\ldots x_M)
\label{eqpsi}
\end{equation}
When $x_i = x_{i+1} -1 $ or $x_M = x_1 + L -1$
(i.e. for configurations where two consecutive sites are occupied),
 the equation
(\ref{eqpsi}) is in principle modified. Insisting that it remains
valid even in these particular cases, requires that the function
$\psi(x_1,\ldots x_M)$ take values in unphysical regions ($x_{i+1} =
x_i$ or $x_M = x_1 + L$) which satisfy the following conditions
arising from the interaction of particles:
\begin{eqnarray}
(1-\beta)\psi(x_1,\ldots x_1+L-1)
-{\rm e}^\gamma \alpha\psi(x_1-1,\ldots x_1+L-1)=0
\label{con1} \\
\psi(\ldots,x_i, x_i +1,\ldots) -\psi(\ldots,x_i, x_i ,\ldots) =0
\;\;\;\mbox{for}\;\;\;1<i<M\;\;\;
\label{con2} \\
\alpha \psi(x_1,x_1+1 \ldots )-\psi(x_1,x_1 \ldots )
+{\rm e}^{-\gamma}  \beta \psi(x_1+1,x_3, \ldots, x_1+L )=0
\label{con3}
\end{eqnarray}
The Bethe ansatz  consists of writing  the eigenfunction $\psi(x_1,\ldots x_M)$
 as 
\begin{equation}
\psi(x_1,\ldots x_M) = {\rm e}^{\gamma\,x_1} \alpha^{x_1}
  \sum_Q
{\cal A}_{Q(1) \cdots Q(M)}( z_{Q(1)})^{x_1} \cdots (z_{Q(M)})^{x_M}\;.
\label{psi}
\end{equation}
where the sum is over all permutations $Q$ of $1 \ldots M$.
The amplitudes ${\cal A}_{Q(1) \cdots Q(M)}$
and  the wave numbers $z_1 \ldots z_M$ are {\it a priori} arbitrary complex numbers.
This ansatz inserted into (\ref{eqpsi})
 gives for the eigenvalue 
\begin{equation}
\lambda ( \gamma) = -(M-1+\alpha) + \sum_{k=1}^{M} \frac{1}{z_k}.
\label{lambdaz}
\end{equation} for any choice of the amplitudes ${\cal A}_{Q(1) \cdots Q(M)}$
and   of the wave numbers $z_1 \ldots z_M$.
For (\ref{psi}) to  fulfil
conditions (\ref{con1}--\ref{con3}), the amplitudes and the wave numbers have to satisfy  
\begin{eqnarray}
{\cal A}_{j\cdots i}&=& (-) \frac{z_j^L}{z_i^L}\ \frac{\left[
(1-\beta)z_i -1\right]}{\left[ (1-\beta)z_j -1\right]} \ {\cal
A}_{i\cdots j}
\label{A1}\\
{\cal A}_{\cdots ji\cdots}&=& (-) \frac{z_j-1}{z_i-1}\ {\cal A}_{\cdots ij\cdots}
\label{A2}\\
{\cal A}_{ji\cdots}&=& (-)\frac{1}{(\alpha z_i-1)}
\left[ (\alpha z_j -1) {\cal A}_{ij\cdots }
	+\alpha \beta z_i z_j^L {\cal A}_{i\cdots j}
	+ \alpha \beta z_j z_i^L  {\cal A}_{j\cdots i} \right]\;.
\label{A3}
\end{eqnarray}
Using (\ref{A1},\ref{A2}) allows (\ref{A3}) to be written as
\begin{equation}
{\cal A}_{ji\cdots}= (-)\frac{(\alpha z_j-1)}{(\alpha z_i-1)}
\left[ 1+ \alpha \beta \frac{z_j^L}{(\alpha z_j-1)\ (bz_j-1)\
(z_j-1)^{M-1}} \frac{z_j-z_i}{z_i-1} \prod_{k=1}^{M} (1-z_k)
\right] {\cal A}_{ij\cdots}
\label{A4}
\end{equation}
where
\begin{equation}
b=(1-\beta)\;.
\end{equation}
Using (\ref{A4})  twice in succession yields, after some algebra,
the following condition on the $z_i$
\begin{eqnarray}
\lefteqn{ \left[
\frac{(\alpha z_i-1)\ (bz_i-1)\
(z_i-1)^{M-1}}{z_i^L} -\alpha \beta  \prod_{k=1}^{M} (1-z_k)
\right] \frac{1}{1-z_i} }
\nonumber \\
&&=
\left[ \frac{(\alpha z_j-1)\ (bz_j-1)\
(z_j-1)^{M-1}}{z_j^L} -\alpha \beta  \prod_{k=1}^{M} (1-z_k)
\right] \frac{1}{1-z_j}\;.
\label{z1}
\end{eqnarray}
The wavefunction $\psi(x_1, \dots x_M)$ corresponding to the largest eigenvalue
$\lambda(\gamma)$ is invariant under translation, therefore
\begin{equation}
{\rm e}^\gamma \alpha \prod_{k=1}^{M} z_k = 1\;.
\label{z2}
\end{equation}

One can rewrite the Bethe equations (\ref{z1})  in terms
of two constants $C,E$ as follows
\begin{eqnarray}
C &=& (-)^{M+1}\alpha \beta \prod_{k=1}^{M}(z_k-1)
\label{C} \\
E &=& -\frac{1}{z_i-1}\ \left[ \frac{(\alpha z_i-1)\ (bz_i-1)\
(z_i-1)^{M-1}}{z_i^L\ C} +1 \right]
\label{E}
\end{eqnarray}
Under this form the Bethe equations (\ref{z1}) are much easier to solve.  One
first finds the $z_i$ solutions of (\ref{E}) (which depend on the
unknown constants $C$ and $E$). Then by inserting these solutions into
(\ref{z2}) and (\ref{C}), the constants $C$ and $E$ are determined.
In  Appendix A, we show (\ref{lambdaA},\ref{alphaA},\ref{EeqA}) that,
in so doing, the eigenvalue $\lambda(\gamma)$ can be written as
\begin{eqnarray}
\lambda( \gamma)= - \sum_{n=1}^\infty {C^n \over n}  \left[  
\oint_1 + \oint_{1 \over \alpha} \right]\! {dz \over 2 \pi i}\ {1 \over z^2} [Q(z)]^n
\label{lambda}
\\
\gamma= - \sum_{n=1}^\infty {C^n \over n} 
\left[  \oint_1 + \oint_{1 \over \alpha} \right]\! {dz \over 2 \pi i}\ {1 \over z} [Q(z)]^n
\label{alpha} \end{eqnarray}
where 
\begin{equation}
Q(z) = {-z^L [1+(z-1)E] \over (bz-1)(\alpha z -1) (z-1)^{M-1}}
\label{Qdef}
\end{equation}
and the constant $E$ is fixed by imposing
 \begin{eqnarray}
0 =  \sum_{n=1}^\infty {C^n \over n} 
\left[  \oint_1 + \oint_{1 \over \alpha} \right]\! {dz \over 2 \pi i}\ {1 \over z-1} [Q(z)]^n \;.
\label{Eeq}
\end{eqnarray}
As is shown in the appendix A the contours of integration in
(\ref{lambda},\ref{alpha},\ref{Eeq}) are small contours which surround
$1$ and $1/\alpha$ but do not surround $1/b\,$; the particular cases
where $\alpha =1$, $b=1$ or $\alpha= b$ can be obtained easily as
limiting cases since all the integrals which appear in the right hand
side of (\ref{lambda},\ref{alpha},\ref{Eeq}) are rational functions of
$\alpha$ and $\beta$.

 Equations (\ref{lambda},\ref{alpha},\ref{Qdef},\ref{Eeq}) determine the exact expression of $\lambda(\gamma)$ for arbitrary $L,M,\alpha$ and $\beta$.
The difference between  these  equations 
and the corresponding equations of \cite{DL} {\bf is}  that in the present case
we have an additional unknown constant $E$. This feature emerges
from the structure of the Bethe equations (\ref{z1}).

\nsection{Exact Expressions for the Velocity and Diffusion Constant}

In principle, one can use (\ref{Eeq}) to expand $E$ in powers of
$C$.  Replacing $E$ by its expansion in powers of $C$ in
(\ref{lambda},\ref{alpha}) gives the expansions of $\lambda$ and
$\gamma$ in powers of $C$. Then one can eliminate $C$ between the two
expansions and this gives $\lambda$ in powers of $\gamma$.  This is
what is done in this section to obtain exact expressions of the
velocity and of the diffusion constant.

 For example to obtain the velocity $v$, one can note from (\ref{alpha})
that $C$  vanishes linearly with  $\gamma$ so that from (\ref{Eeq}), the limiting value $E(0)$ of $E$ at $\gamma=0$ is
\begin{equation}
E(0) = -\frac{X_{L,M}}{X_{L,M-1}}
\label{EX}
\end{equation}
where $X_{L,M}$ is defined by
\begin{equation}
X_{L,M}=  
\left[ \oint_1 + \oint_{1 \over \alpha} \right] {dz \over 2 \pi i}
\frac{z^L}{(z-1)^M} \frac{1}{(bz-1)(\alpha z-1)}\;.
\label{Xdef}
\end{equation}
Then from (\ref{lambda},\ref{alpha}), one finds that for $\gamma$ small
$\lambda(\gamma) = v \gamma + O(\gamma^2) $ with
 the velocity $v$ given by
\begin{equation}
v= \frac{X_{L,M}X_{L-2,M-2}- X_{L,M-1}X_{L-2,M-1} }{Z_{L,M}}
\label{vX}
\end{equation}
where
\begin{equation}
Z_{L,M} = X_{L,M}X_{L-1,M-2} -X_{L,M-1} X_{L-1,M-1}\; .
\label{Zdef}
\end{equation}
Expression (\ref{vX}) may be simplified by using
\begin{equation}
X_{L,M}=X_{L-1,M}+X_{L-1,M-1}
\label{Xsimp}
\end{equation}
to obtain
\begin{equation}
v= \frac{Z_{L-1,M}- Z_{L-1,M-1}}{Z_{L,M}}\;.
\label{vZ}
\end{equation}
In a similar fashion one obtains   from the second derivatives of
(\ref{lambda},\ref{alpha},\ref{Eeq}) at $\gamma=0$,
 after a good deal of straightforward but tedious algebra,
\begin{eqnarray}
\Delta &= &\frac{X_{L,M-1}}{ Z_{L,M}^2}
\left\{ W_{2L,2M-1} X_{L-2,M-2} -W_{2L-2,2M-2} X_{L,M-1}
\nonumber \right. \\
&&\left. \hspace{1.0in} + v \left[\, W_{2L-1,2M-2} X_{L,M-1} -W_{2L,2M-1} X_{L-1,M-2} \,\right]
\right\}
\label{Deltaeq1}
\end{eqnarray}
where $W_{2L,2M}$ is defined by
\begin{equation}
W_{2L,2M}= 
\left[  \oint_1 + \oint_{1 \over \alpha} \right] \frac{dz}{2\pi i}
\frac{z^{2L}}{(z-1)^{2M}} \frac{[1+E(0)(z-1)]^2}{(bz-1)^2(\alpha z-1)^2}\;.
\label{Wdef}
\end{equation}
Expression (\ref{Deltaeq1})
can be simplified by using (\ref{vX}--\ref{Xsimp}) to obtain
\begin{eqnarray}
\Delta &=& \frac{X_{L,M-1}^2}{ Z_{L,M}^3}
\left\{ W_{2L,2M-1} Z_{L-1,M-1} -W_{2L-2,2M-2} Z_{L,M}
\nonumber \right. \\
&&\left. \hspace{1.0in} +   W_{2L-1,2M-2}\left[\,
Z_{L-1,M} -Z_{L-1,M-1} \,\right]\,
\right\}\;.
\label{Deltaeq}
\end{eqnarray}
 Alternatively one could obtain (\ref{Deltaeq})
directly from (\ref{lambda}-\ref{Eeq}) by using, for example,
 Mathematica.
The integrals in the above expressions can be evaluated by
residues. In this way one can show that the integral expressions for
$Z_{L,M}$ and $v$ given by (\ref{Xdef},\ref{Zdef},\ref{vZ}) are equivalent to
those derived in \cite{BerChine,Kirone} within the  matrix product
formulation.  However exact evaluation of the integrals involved in
the diffusion constant (\ref{Deltaeq}) results, in general, in
cumbersome expressions.

\noindent {\bf Remark:}
For the case of a second class particle ($\alpha=\beta=1$)
simplification of (\ref{Deltaeq}) is possible and one recovers the
expression first presented in \cite{DECam}:
\begin{eqnarray}
\Delta &=& 2 \frac{(2L-3)!}{(2M-1)!(2L-2M+1)!} 
\left[ \frac{(M-1)!(L-M)!}{(L-1)!} \right]^2
\nonumber\\
&&\times \left[ (L-5)(M-1)(L-M) +(L-1)(2L-1)\right].
\label{Delta2c}
\end{eqnarray}
However the derivation of (\ref{Delta2c}) from
(\ref{Deltaeq}) is tedious and not illuminating therefore we do not
present it.
\nsection{Asymptotics and Phase Diagram}
In order to obtain the phase diagram it suffices
to determine the asymptotic form of $Z_{L,M}$ which in turn
determines the asymptotic form of the velocity (\ref{vZ}).
We carry this out in detail in appendix B, but would like to outline
here how the different phases arise. We restrict ourselves to
$\beta<1$ ($b>0$) and $\alpha <1$.

Consider the quantity $X_{L,M}$ given by
(\ref{Xdef}). In the limit of $L,M$ large (with fixed $\rho=L/M$) there are three
possible dominant contributions to the integral, all lying on the real axis:
a saddle point   
at $z_c = 1/(1-\rho)$; a pole at $z=1/b$ and a pole
at $z=1/\alpha$. The possible dominant and subdominant contributions
to $X_{L,M}$ gives rise to four phases  as follows.

\begin{enumerate}
\item
If $1/\alpha < z_c  < 1/b$ the contours of the two
integrals in (\ref{Xdef})
may be merged and deformed to pass through the saddle point.
Therefore only the saddle point contributes.

\item
If $1/\alpha < z_c$ and $1/b < z_c$ the contours
of the two integrals may be merged and deformed to pass through the
saddle point. However the contour must
make a clockwise detour around
the pole at $z=1/b$. Therefore the pole at $z=1/b$ is the dominant
contribution and the
saddle point is the subdominant contribution.

\item
If $1/\alpha > z_c$ and $1/b > z_c$ 
the two integrals give separate contributions.
The pole at $ z=1/\alpha$  gives the dominant
contribution and the integral around $z=1$ may be deformed
to pass through the saddle point
and gives the subdominant contribution.

\item
If $1/\alpha > z_c > 1/b$
the two integrals give separate contributions:
the pole at $ z=1/\alpha$ and the clockwise integral
around the pole at $z=1/b$.
\end{enumerate}

In appendix B the dominant contributions to the desired integrals are
evaluated and expressions (\ref{res5}--\ref{res3}) are established.
At this point we can already see the interesting feature that since
$Z$ in (\ref{Zdef}) is a difference of products of $X$, the
subdominant contribution as well as the dominant contribution to the
integral $X$ must be evaluated to obtain $Z$ and the velocity.  Also
note that in phase where $1/\alpha < z_c < 1/b\,$, power law decays in
correlation functions, for example the density profile, will arise
from the saddle point being dominant. In the other phases a dominant
pole will give rise to exponential decays.  Of particular interest is
the phase where $1/\alpha > z_c > 1/b$ and the two poles compete.  As
described in the introduction this is the phase where a shock exists.

\nsection{Scaling of the Large Deviation Function for
$1 - \alpha <    \rho < \beta$}

In this phase, because the integrals are dominated by the saddle
point, the analysis of the asymptotics of the exact expression of
$\lambda(\gamma)$  given by (\ref{lambda}--\ref{Eeq}) is rather
different from the other cases.  In these expressions a contour
integral actually implies two integrals around $z=1$ and $z=1/\alpha$
but in this phase, for large $L$, one expects all the integrals to be
dominated by their saddle point $z_c=1/ (1-\rho)$.  So to lighten the
notation we write a single integral.  Let us replace the variables $z$
and $E$ in (\ref{lambda}--\ref{Eeq}) by $y$ and $F$
\begin{equation}
z= z_c + y
\label{ydef}
\end{equation}
and
\begin{equation}
E= - {1 \over z_c-1} + {1 \over z_c-1}{F \over L} \;.
\label{Edef}
\end{equation}
Clearly, the values of $y$ which contribute  to the integrals in 
(\ref{lambda},\ref{alpha},\ref{Eeq}) are of order $y= O(L^{-1/2}) $
so that one can rewrite (\ref{lambda},\ref{alpha},\ref{Eeq}) as
\begin{eqnarray}
\lambda( \gamma)&=&
 -{1\over z_c^2} S_0 + {2\over z_c^3} S_1 - {3\over z_c^4} S_2
 + {4\over z_c^5} S_3 + O\left( S_3 \over L^{1/2} \right)
 \label{lambda3} \\
 \gamma &=& -{1\over z_c} S_0 + {1\over z_c^2} S_1 - {1\over z_c^3} S_2
 + {1\over z_c^4} S_3  + O\left( S_3 \over L^{1/2} \right) \label{gamma3}\\
 0 &=& {1\over z_c -1} S_0 - {1 \over (z_c-1)^2} S_1 + {1\over (z_c-1)^3}
S_2 - {1\over (z_c-1)^4} S_3 + O\left( S_3 \over L^{1/2} \right)  \label{E3}
\end{eqnarray}
 where $S_p$ is given by
\begin{equation}
S_p = \sum_{n \geq 1} {C^n \over n}  \oint {dy \over 2 \pi i}
\left[ R(y)
 \left( {-y \over z_c-1} +{F\over L} + { F y \over L (z_c -1)} \right) \right]^n
\ \  y^p
\label{Spdef}
\end{equation}
 with
\begin{equation}
R(y)= {1 \over (1-b z_c - by ) ( \alpha z_c + \alpha y -1)} {(z_c +
y)^L \over (z_c + y-1)^{M-1} }\;.
\end{equation}
Under the assumption (which we will check later)  that $F$ is of order 1  (in the large $L$ limit), 
if we define
\begin{equation}
g(y) = \log (z_c+y) - \rho \log(z_c -1 +y) 
\end{equation}
we can evaluate the leading orders of the integrals (\ref{Spdef}).

\noindent \underline{For $p$ odd}, the leading large $L$ behaviour is  given by
\begin{equation}
S_p \simeq  { (-)^{p+3 \over 2} \over \sqrt{2 \pi}}  {1- \rho \over \rho} {D \over L^{1+{p \over 2}}} \left({1 \over g''(0)} \right)^{1+{p \over 2 }} p!! 
\label{Spodd}
\end{equation}
where 
\begin{equation}
D= C e^{L g(0)} {1 - z_c \over (b z_c -1)  ( \alpha z_c -1) }
\label{Ddef}
\end{equation}

\noindent \underline{For $p$ even} the leading order in the range where $D$ is of order 1 is 
\begin{eqnarray}
S_p &\simeq&  { (-)^{p \over 2} \over \sqrt{2 \pi}}   {D \over L^{p+3 \over 2}}
\left({1 \over g''(0)} \right)^{p+5 \over 2 }
 \nonumber \\
&&\times \left\{   F  g''(0)^2 (p-1)!! 
  +{(1-\rho)^2\over \rho} 
\left[ {1 \over \rho} - {b\over \rho +b -1}
    - {\alpha \over \alpha -1 +\rho} \right] g''(0) (p+1)!!  
\right.
\nonumber \\
&&\hspace{1.0in}
\left.  -{1-\rho \over \rho} {g'''(0)\over 6} (p+3) !!   \right\}
\nonumber \\
&&
 + {D^2 \over 2} \left({1-\rho \over \rho}\right)^2 {1 \over L^{p+3 \over 2}} \left(1 \over 2 g''(0)\right)^{p+3 \over 2} {(-)^{p +2\over 2}  \over \sqrt{2 \pi}} (p+1)!! 
\label{Speven}
\end{eqnarray}
where we define $(p-1)!! =1$ for $p=0$.
$F$ is fixed through (\ref{E3}) by
\begin{equation}
S_1 \simeq (z_c-1) S_0
\label{E4}
\end{equation}
and from (\ref{lambda3},\ref{gamma3}) we obtain
\begin{eqnarray}
\gamma &\simeq& -{1 \over z_c^2 (z_c-1)} S_1 \label{gamma4}\\
\lambda(\gamma)- \gamma {2 - z_c \over z_c}
&\simeq& {1 \over z_c^5 (z_c-1)^2} [ S_3 ( 2- 3 z_c) - S_2 z_c (1- z_c) ] \;.
\label{lambda4}
\end{eqnarray}
Using (\ref{Spodd}) and (\ref{Speven}),  (\ref{E4}) gives 
\begin{equation}
F \simeq {1 \over (1-\rho)}
\left[ -\frac{1+\rho}{\rho} + 
\frac{b}{b-1+\rho}+\frac{\alpha}{\alpha-1+\rho} \right]
 +{D \over  {\rho(1-\rho)2^{5/2}} }
\end{equation}
which is consistent with our earlier  asumption that $F$ is of order 1.
  From (\ref{gamma4}),(\ref{lambda4}) we find
\begin{eqnarray}
\gamma &=& {-D \over \sqrt{2 \pi \rho (1- \rho)} L^{3/2}} \\
\lambda(\gamma) - \gamma (1- 2 \rho) &\simeq& \frac{2 \gamma }{L}
\left[ 2-\rho \left(1 + \frac{1-\beta}{\rho-\beta}
 + \frac{\alpha}{\alpha-1+\rho} \right)
\right]
 + \gamma^2 L^{1/2} {\sqrt{\pi \rho (1- \rho)} \over 8} 
\end{eqnarray}
as announced in (\ref{res6}).

\nsection{Discussion}

In this work we have shown how the Bethe ansatz can be used to
 calculate exactly via (\ref{lambda}--\ref{Eeq}) the large deviation
 function of the displacement of the defect particle. By analyzing the
 asymptotics of (\ref{lambda}--\ref{Eeq}) we could recover the
 velocity and phase diagram \cite{Kirone,BerChine} of the asymmetric
 exclusion model with a moving defect.  The approach also allows new
 results (such as all the cumulants of the displacement of the defect)
 to be obtained, in particular the diffusion constant of the defect
 particle in the various phases. This adds to the body of knowledge
 concerning diffusion constants within the asymmetric exclusion
 process.  For example the exact expression for the diffusion constant
 of a first-class particle calculated in \cite{DEM} allowed the
 determination of a universal amplitude for the centre of mass
 fluctuation for growth processes described by the one dimensional KPZ
 equation \cite{KRUG}. The diffusion constant of a second class
 particle is of interest since it is closely related to the motion of
 shocks.  In \cite{FF} the diffusion constant for a second class
 particle starting at the origin of an infinite lattice with a shock
 initial condition was calculated. Our results (\ref{res3}) for the
 phase exhibiting a shock ($\beta < \rho < 1-\alpha$) exactly agrees
 with that of \cite{FF}. This is of interest since it shows that a
 single defect can provoke a shock in the ring geometry with the same
 behaviour as a shock on the infinite line \cite{DGLS}.  If the
 fluctuations of the shock on a ring with a defect are identical to
 the fluctuations of the shock on an infinite line, this means that
 our results (\ref{lambda}-\ref{Eeq}) should give the whole large
 deviation function of a shock position on an infinite line.  The
 behaviour of shocks and shock fluctuations is also connected to the
 phase diagram and density profile for systems with open boundary
 conditions \cite{KSKS}.

 When $\gamma \to 0$ the wave
function  (\ref{psi}) reduces to the steady state
probabilities for each configuration. Therefore, in principle, the
steady state of the system, previously constructed using a matrix
product \cite{DJLS,BerChine,Kirone}, can be extracted from the present Bethe ansatz
$(\ref{psi})$ by taking the limit $\gamma \to 0$.  
This shows that a
co-ordinate Bethe ansatz is capable of describing non-trivial steady
states of stochastic systems. 
In particular it would be interesting to
understand further how this works and to see if 
the approach might be  generalisable to larger numbers of
species.  The matrix product steady state of the present model is very
closely related to that of the ASEP with open boundary conditions
\cite{DEHP,DJLS}.  It would be of great interest to determine whether some
implementation of the
Bethe ansatz, perhaps related to that of the present work, could be
used to recover the steady state with  open boundary conditions. 
A major difficulty in doing so is that
the particle number is
not conserved with open boundaries.

Returning to the case of a second class particle it is of interest to
review how its dynamics are related to the spreading of excess
mass. The central idea, termed coupling \cite{Liggett}, is well-known
in the mathematical community but less so within physics. Consider two
systems containing only first class particles, identical except that
one system has $M$ particles and the other $M-1$ particles.  The two
systems start from initial conditions differing only by the position
of the extra particle in the system with $M$ particles. In order to
implement the dynamics one can consider at each time step randomly
choosing a pair of sites $i,i+1$ to update; then if there is particle
at site $i$ and a hole at site $i+1$ the particle is moved forward.
In the dynamics let us choose the same pairs of sites in the two systems
at each update (one can think of using the same random numbers in a
Monte Carlo program). Now it is easy to convince oneself that after
any length of time the configurations of the two systems will differ
only by the position of the extra particle (note that if we label the
particles, the label of the extra particle will change under the
dynamics).  Further, one can convince oneself that the position of the
extra particle has precisely the dynamics of a second class particle
in the ASEP. Conversely, the system comprising $M-1$ first class
particles and one second class particle that we have studied describes
the motion of an extra particle added to a system of $M-1$ particles.
Therefore the diffusion constant of the second class particle we have
calculated here serves to describe the spreading of excess mass in the
ASEP.

Approximate calculations such as mode coupling \cite{vBKS,vB} have led
to the following understanding of the motion of excess mass
fluctuations: the drift speed is $1-2\rho$ as in (\ref{res5}) for
$\alpha=\beta=1$ and the spreading of density fluctutations around the
drift grows as $t^{2/3}$ on an infinite system {\it i.e.}  it is
superdiffusive.  This superdiffusive motion can be recovered from the
$L^{1/2}$ finite system size dependence (\ref{res5}) of the diffusion
constant of a second class particle \cite{DECam} if we assume that a
scaling form holds and the variance of the distance travelled by the
second class particle can be written as
\begin{equation}
\langle y_t^2 \rangle - \langle y_t \rangle^2  \sim t L^{1/2} f(t/L^z)
\end{equation}
where $z$ is the dynamic exponent and $f(x)$ is a scaling function
tending to a constant as $x \to \infty$. Now the dynamic exponent for
the ASEP is known by the Bethe ansatz to be $z=3/2$ \cite{GS,Kim} and
we expect the same exponent in the present model.
  In the
limit $L \to \infty$ for large but fixed $t$, the variance $\langle
y_t^2 \rangle - \langle y_t \rangle^2$ should not depend on system
size therefore the scaling function must obey $f(x) \sim x^{1/2z}$ as
$x \to 0$. We then find in this infinite system limit that $\langle
y_t^2 \rangle - \langle y_t \rangle^2 \sim t^{4/3}$ so that the
typical spread of density fluctuations grows as $t^{2/3}$.  The
spreading of mass fluctuations is also related to the scaling length
$\xi \sim t^{2/3}$ of the KPZ equation in one dimension  (see \cite{KRUG} for detailed
discussion).

Finally let us mention that one can easily extend the calculation of
the present paper to calculate the joint distribution of the distance
$y_t$ covered by the defect particle and of the total distance $Y_t$
covered by all the first class particles.  One can show that
\begin{equation}
\lambda(\gamma, \delta) = \lim_{t \to \infty} { \ln \left[\langle
e^{\gamma y_t +\delta Y_t} \rangle\right] \over t}
\end{equation}
is still given by (\ref{lambdaz}) for arbitrary $L$ and $M$ where the
 $z_i$ and the constants $C$ and $E$ satisfy
\begin{equation}
{\rm e}^{\gamma + (M-1) \delta} \  \alpha \prod_{k=1}^{M} z_k = 1\;.
\end{equation}
\begin{eqnarray}
C &=& (-)^{M+1} \ \alpha \beta \  e^{\delta L}  \  \prod_{k=1}^{M}(z_k-1)
\end{eqnarray}
instead of (\ref{z2},\ref{C}) with (\ref{E}) unchanged.

\noindent {\bf Acknowledgements}

\noindent MRE is a Royal Society University Research Fellow
and thanks Laboratoire de Physique Statistique de l'ENS
for hospitality during a visit when this work was in progress.

\appendix{A: Analysis of the Bethe ansatz equations}
\setcounter{equation}{0}
\def\theequation{A\arabic{equation}}
 The  solution for $\{ z_i \}$ of (\ref{z2}--\ref{E}) which gives $\lambda \to 0$ as $\gamma \to 0$
is of the form $z_1 \to 1/\alpha$ and $z_k \to 1$ for
$2 \leq k \leq M $.
We analyse separately the cases $\alpha \neq 1$ and
$\alpha = 1$
and for simplicity assume $\beta \neq 1$ and $\beta \neq \alpha$
although there is no problem in extending the analysis to include
these cases.

\noindent{ \underline{{\it Case} $\alpha \neq 1$}}

\noindent
Consider the root $z_1$ and the $M-1$ roots  $z_k$ for $2 \leq k \leq M $ of
(\ref{E}) which we rewrite as
\begin{equation}
 (z-1)^{M-1} (bz-1) (\alpha z-1) + z^L C [1 + (z-1) E ] =0 
\label{poly}
\end{equation}
such that $z_1 \to 1/\alpha$ and $z_k \to 1$ as $C \to 0$.
Define
\begin{equation}
R(z) = {- z^L [ 1 + (z-1) E ] \over (bz-1)(\alpha z -1 ) } \;,
\end{equation}
then if $z_k$ is a root of (\ref{poly})
such that $z_k \to 1$ as $C \to 0$ one has for small $C$ 
\begin{equation}
z_k = 1 + \left[ C e^{2 i \pi k} R(z_k) \right]^{1 \over M-1}\;.
\label{zk}
\end{equation}
We wish to calculate expressions (\ref{lambdaz},\ref{z2},\ref{C})  of the form $\sum_k h(z_k)$
(e.g. Eq. \ref{lambdaz} where $h=1/z$).
If $h(z)$ is analytic near $z=1$,
one has by the residue theorem
\begin{equation}
h(z_k) =  \oint_1 {dz \over 2 \pi i}\, h(z) 
{1 - \left[ C e^{2 i \pi k} R(z_k) \right]^{1 \over M-1} R'(z)/ [(M-1) R(z) ] \over
z- 1 - \left[ C e^{2 i \pi k} R(z_k) \right]^{1 \over M-1} } 
\end{equation}
where the contour is a circle centred on 1 and of  radius $\epsilon$
with $ |C|^{1\over M-1} \ll \epsilon \ll 1$.
Expanding in powers of $C^{1\over M-1}$ and after an integration by parts, this gives
\begin{equation}
h(z_k) = h(1) + \sum_{p=1}^\infty {1 \over p} \left[ C
e^{2 i \pi k } \right]^{ p \over M-1} \oint_1 {dz \over 2 \pi i}\, h'(z) {
[R(z)]^{p\over M-1} \over (z-1)^p}
\end{equation}
Then summing over the roots $2 \leq k \leq M$ leads to
\begin{equation}
\sum_{k=2}^{M} h(z_k) = (M-1) h(1) + \sum_{n=1}^\infty {1 \over n}   C^n   
\oint_1 {dz \over 2 \pi i}\, h'(z) { [R(z)]^n \over (z-1)^{n(M-1)}} 
\label{hk}
\end{equation}
Similarly, if  $h(z)$ is analytic near $z= 1/\alpha$ and we define
\begin{equation}
S(z) = {- z^L [ 1 + (z-1) E ] \over  \alpha (bz-1)( z -1 )^{M-1} } 
\end{equation}
then
\begin{equation}
h(z_1) =  \oint_{1/\alpha} {dz \over 2 \pi i}\, h(z) {1 - C S'(z) \over z - {1 \over \alpha} - C S(z) } 
\end{equation}
where the contour is a circle centred on $1/ \alpha$  and of  radius $\epsilon$
with $ |C| \ll \epsilon \ll 1$.
After expanding in powers of $C$ and an integration by parts  this gives
\begin{equation} 
h(z_1) = h \left( {1 \over \alpha } \right) + \sum_{n=1}^\infty {C^n \over n} 
\oint_{1 \over \alpha}  {dz \over 2 \pi i}\, h'(z) {[S(z)]^n \over (z- {1 \over \alpha})^n} 
\label{h1}
\end{equation}
Therefore, if one defines $Q(z)$ by
\begin{equation}
Q(z) = {-z^L [1+(z-1)E] \over (bz-1)(\alpha z -1) (z-1)^{M-1}}
  = {R(z) \over (z-1)^{M-1}} = {S(z) \over z- {1\over \alpha}}
\label{QdefA}
\end{equation}
one finds  by combining (\ref{h1},\ref{hk})
\begin{equation}
\sum_{k=1}^{M} h(z_k)= (M-1) h(1) + h \left( {1 \over \alpha } \right) 
+ \sum_{n=1}^\infty {C^n \over n}  \left[  \oint_1 + \oint_{1 \over \alpha} \right] {dz \over 2 \pi i}\, h'(z) [Q(z)]^n  \;.
\label{heq}
\end{equation}

Writing (\ref{lambdaz}),(\ref{z2}) as
\begin{equation}
\lambda(\gamma) =- (M-1+ \alpha) + \sum_{k=1}^{M}  {1 \over z_k} 
\end{equation}
\begin{equation}
\gamma = - \log \alpha - \sum_{k=1}^{M} \log z_k 
\end{equation}
one finds that
\begin{eqnarray}
\lambda( \gamma)= - \sum_{n=1}^\infty {C^n \over n} 
\left[ \oint_1 + \oint_{1 \over \alpha} \right]\! {dz \over 2 \pi i}\, {1 \over z^2}
[Q(z)]^n
\label{lambdaA}
\\
\gamma= - \sum_{n=1}^\infty {C^n \over n} 
\left[  \oint_1 + \oint_{1 \over \alpha} \right]\! 
{dz \over 2 \pi i}\, {1 \over z} [Q(z)]^n  
\label{alphaA} 
\end{eqnarray}
where $Q(z)$ is given by (\ref{QdefA}).
Then,  with the use of (\ref{zk}),  replacing (\ref{C}) by
\begin{equation}
(1- z_1) \prod_{k=2}^{M} \left[ R(z_k) \right]^{1 \over M-1}
 = {1 \over \alpha \beta}
\label{constraint}
\end{equation}
and using the fact (\ref{hk},\ref{h1}) that
\begin{eqnarray}
0 &=& \ln(\alpha \beta) + \ln(1-z_1) + \frac{1}{M-1}
\sum_{k=2}^{M} \ln R(z_k) \\ &=&
\sum_{n=1}^\infty {C^n \over n} {1 \over 2 \pi i} \left[ {1 \over M-1}
\oint_1 dz\, {R'(z) [R(z)]^{n-1} \over (z-1)^{n(M-1)}} + \oint_{1 \over
\alpha} dz\, {1 \over z-1} { [S(z)]^{n} \over (z-{1 \over \alpha})^{n}}
\right]
\end{eqnarray}
one finds that (\ref{C}) is satisfied if
\begin{eqnarray}
0 =  \sum_{n=1}^\infty {C^n \over n} 
\left[  \oint_1 + \oint_{1 \over \alpha} \right]\!
{dz \over 2 \pi i}\, {1 \over z-1} [Q(z)]^n  
\label{EeqA}
\end{eqnarray}

\noindent{ \underline{ {\it Case} $\alpha =1$}}

\noindent Let \begin{equation}P(z)= {-z^L [1+(z-1)E] \over (bz-1)}
\end{equation} Then if $z_k$ is the root such that $z_k \to 1$ as $C
\to 0$ with for small $C$
\begin{equation}z_k = 1 + \left[ C e^{2 i \pi k} P(z_k) \right]^{1 \over M}
\end{equation}
and if $h(z)$ is analytic near $z=1$, one has
\begin{equation}h(z_k) = \oint_1 {dz \over 2 \pi i} \, h(z) {1 - 
\left[ C e^{2 i \pi k} P(z_k) \right]^{1 \over M}P'(z)/ [M P(z) ] \over
z- 1 - \left[ C e^{2 i \pi k} P(z_k) \right]^{1 \over M} }
\end{equation}

Then summing over the roots $0 \leq k \leq M-1$ leads to
\begin{equation} \sum_{k=0}^{M-1} h(z_k) = M h(1) + \sum_{n=1}^\infty {1 \over n}
C^n \oint_1 {dz \over 2 \pi i}\, h'(z) { [P(z)]^n \over (z-1)^{n M}}
\end{equation} Therefore the equations for case $\alpha=1$ are given by exactly
the same expressions as the case $\alpha \neq 1$
(\ref{lambda},\ref{alpha},\ref{Eeq}) with the replacement
$$ \left[  \oint_1 + \oint_{1 \over \alpha} \right]  \to
  \oint_1 \;. $$.
\label{appbethe}

\appendix{B: Asymptotic  Evaluation of Velocity and Diffusion Constant}
\setcounter{equation}{0}
\def\theequation{B\arabic{equation}}
\vspace{-0.5in}
\nsubsecnn{Evaluation of the Velocity}
\noindent \underline{For $1/\alpha < z_c$ and $1/b > z_c$ }\\
\noindent In this phase (\ref{Xdef}) is dominated by
 $S_{L,M}$  the saddle point contribution
\begin{equation}
X_{L,M} = S_{L,M} +  O(S_{L,M}/L)
\label{XS}
\end{equation}
where
\begin{eqnarray}
S_{L,M}& =& \frac{1}{\sqrt{2\pi L}} 
\frac{ [\rho(1-\rho)]^{1/2}}{(b+\rho-1)(\alpha+\rho-1)}
 \frac{z_c^L}{ (z_c-1)^M } \;.
\label{Sdef}
\end{eqnarray}
However the leading contributions to $X_{L,M}$
cancel in (\ref{Zdef}). To evaluate the next leading contribution
we write $Z_{L,M}$ as a double integral using $(\ref{Xdef})$
\begin{eqnarray}
Z_{L,M}= \oint  \frac{dz}{2\pi i}\frac{1}{(bz-1)(\alpha z-1)}
\oint  \frac{d\tilde{z}}{2\pi i}\frac{1}{(b\tilde{z}-1)(\alpha \tilde{z}-1)}
\frac{z^L}{(z-1)^{M}} 
\frac{\tilde{z}^L}{(\tilde{z}-1)^{M}} 
\frac{ (\tilde{z}-1)(\tilde{z}-z)}{\tilde{z}}
\label{Zintdef}
\end{eqnarray}
The double integral is dominated by the saddle point
$z=\tilde{z}=z_c=1/(1-\rho)$ and can be evaluated to be
\begin{eqnarray}
Z_{L,M}&\simeq&  \frac{\rho S_{L,M}^2}{(1-\rho) L}
\end{eqnarray}
 From (\ref{vZ}) one obtains $v=1/z_c - (z_c-1)/z_c$ so that
\begin{equation}
v= 1- 2\rho 
\end{equation}
as in (\ref{res5}). 

\noindent \underline{For $1/\alpha < z_c$ and $1/b < z_c$} \\
\noindent As explained in section 6, in this phase the dominant contribution to
(\ref{Xdef}) is the  pole at $1/b$ and the saddle point is subdominant:
\begin{equation}
X_{L,M} = B_{L,M} + S_{L,M}  +O(S_{L,M}/L)
\label{XBS}
\end{equation}
where  $B$ represents the contribution of the clockwise contour around
the pole $z=1/b$ 
\begin{eqnarray}
B_{L,M}&=& \frac{1}{b-\alpha}  \frac{1}{(1-b)^M\ b^{L-M}}\;.
\label{Bdef}
\end{eqnarray}
Therefore, from (\ref{Zdef})
\begin{equation}
Z_{L,M} \simeq \frac{(1-bz_c)^2}{bz_c} B_{L,M}S_{L,M}
\label{ZBS}
\end{equation}
and from (\ref{vZ})
\begin{equation}
v= 1-\beta -\rho
\label{vBS}
\end{equation}
as in (\ref{res2}).

\noindent \underline{For $1/\alpha > z_c$ and $1/b > z_c$}\\
In this phase the pole at $1/\alpha$ is the dominant contribution
to (\ref{Xdef}) and the saddle point is  subdominant
\begin{equation}
X_{L,M} = A_{L,M} + S_{L,M} + O(S_{L,M}/L)
\label{XAS}
\end{equation}
where  $A_{L,M}$ is given by
\begin{eqnarray}
A_{L,M}&=& \frac{1}{b-\alpha}  \frac{1}{(1-\alpha)^M\ \alpha^{L-M}}\;.
\label{Adef}
\end{eqnarray}
$Z_{L,M}$ may be determined by symmetry considerations from the
previous phase:
under interchange of $\alpha$ and $b$ (\ref{XBS}) becomes
(\ref{XAS}) and (\ref{ZBS}) becomes
\begin{equation}
Z_{L,M} \simeq \frac{(1-\alpha z_c)^2}{\alpha z_c} A_{L,M}S_{L,M} 
\label{ZAS}
\end{equation}
and we find
\begin{equation}
v= \alpha-\rho
\label{vAS}
\end{equation}
as in (\ref{res1}).

\noindent \underline{For $1/\alpha > z_c$ and $1/b < z_c$}\\
In this phase
\begin{equation}
X_{L,M} \simeq A_{L,M}+B_{L,M} +O(S_{L,M})
\label{XAB}
\end{equation}
so that
\begin{equation}
Z_{L,M} \simeq \frac{(b-\alpha )^2}{b \alpha} A_{L,M}B_{L,M} 
\label{ZAB}
\end{equation}
and
\begin{equation}
v=\alpha - \beta
\label{vAB}
\end{equation}
as in (\ref{res3}).

\nsubsecnn{Evaluation of Diffusion Constant}
In order to compute the diffusion constant given by (\ref{Deltaeq})
\begin{equation}
\Delta = \frac{X_{L,M-1}^2}{Z_{L,M}^3} U_{L,M}
\label{DeltaU}
\end{equation}
where
\begin{equation}
U_{L,M} = W_{2L,2M-1} Z_{L-1,M-1} +
W_{2L-1,2M-2}(Z_{L-1,M}-Z_{L-1,M-1}) -W_{2L-2,2M-2} Z_{L,M}\;.
\label{Udef}
\end{equation}
we need first to evaluate the asymptotics of $U_{L,M}$.
Using the integral definitions 
(\ref{Wdef},\ref{Zintdef}) we may write 
(\ref{Udef}) as 
\begin{eqnarray}
U_{L,M}&=&\oint \frac{d w}{2\pi i}
\frac{w^{2L}}{(w-1)^{2M}} \frac{[1+E(0)(w-1)]^2}{(bw-1)^2(\alpha w-1)^2}
\oint \frac{dz}{2\pi i}
\frac{z^L}{(z-1)^M} \frac{1}{(bz-1)(\alpha z-1)} 
\nonumber \\
&&\times \oint \frac{d\tilde{z}}{2\pi i}
\frac{\tilde{z}^L}{(\tilde{z}-1)^M} \frac{(\tilde{z}-1)(\tilde{z}-z)}{(b\tilde{z}-1)(\alpha \tilde{z}-1)}
\frac{w-1}{w^2}\frac{w-z}{z}\frac{w-\tilde{z}}{\tilde{z}^2}
\label{Uint}
\end{eqnarray}
\noindent \underline{For $1/\alpha > z_c$ and $1/b > z_c$} \\
\noindent In this phase all integrations are dominated by their saddle points.
However care is required to correctly identify  the first
non-vanishing contribution to (\ref{Uint}). This is done most
systematically by considering the scaling of the large deviation
function and we carry this out in section~7 where we show
\begin{equation}
\Delta \simeq \frac{( L \pi \rho(1-\rho))^{1/2} }{4}\;.
\end{equation}
\noindent \underline{For $1/\alpha < z_c$ and $1/b < z_c$}\\
\noindent 
To evaluate  (\ref{Uint}) we carry out the integrals in sequence.
In the first integral over $\tilde{z}$ we keep an {\it apparently}
subdominant term  (proportional to $S_{L,M}$) as well the term proportional to
$B_{L,M}$, because
when we integrate over $z$ we find both terms give  leading
contributions proportional to $S_{L,M} B_{L,M}$
{\it i.e.} the dominant contributions to the triple integral
come from $w$ at the pole, one of $z,\tilde{z}$ at the pole and the other
at the saddlepoint:
\begin{eqnarray}
U_{L,M}&\simeq&\oint \frac{d w}{2\pi i}
\frac{w^{2L-2}}{(w-1)^{2M-1}} \frac{[1+E(0)(w-1)]^2}{(bw-1)^2(\alpha w-1)^2}
\oint \frac{dz}{2\pi i}
\frac{z^L}{(z-1)^M} \frac{1}{(bz-1)(\alpha z-1)} 
\nonumber
\\
&&\times
\left[
 B_{L,M} \frac{(b-1)(bw-1)(w -z)(bz-1)}{b z}
+
 S_{L,M} \frac{(z_c-1)(w-z_c)(w -z)(z_c-z)}{z_c^2 z}
\right] \nonumber
\\
&\simeq&\oint \frac{d w}{2\pi i}
\frac{w^{2L-2}}{(w-1)^{2M-1}} \frac{[1+E(0)(w-1)]^2}{(bw-1)^2(\alpha w-1)^2}
S_{L,M} B_{L,M}
\nonumber
\\
&&\times
\left[
 \frac{(b-1)(bw-1)(w -z_c)(bz_c-1)}{b z_c}
+
\frac{(z_c-1)(w-z_c)(bw -1)(bz_c-1)}{b z_c^2 }
\right] \nonumber \\
&=&
S_{L,M} B_{L,M} \frac{(bz_c-1)^2}{b z_c^2}\oint \frac{d w}{2\pi i}
\frac{w^{2L-2}}{(w-1)^{2M-1}} 
\frac{[1+E(0)(w-1)]^2\ (w-z_c)}{(bw-1)(\alpha w-1)^2}
\nonumber\\
&\simeq&
S_{L,M}B_{L,M}^3 \frac{ (1-b)(bz_c-1)^3}{z_c^2}
\left[1+E(0)\left({1-b \over b}\right)\right]^2\;.
\end{eqnarray}
In this phase the behaviour (\ref{XBS}) and the form of
(\ref{Bdef},\ref{Sdef})  imply that $E(0)$ given by (\ref{EX}) becomes
\begin{equation}
E(0) \simeq - \frac{b}{1-b} \left[ 1 +\frac{(1-bz_c)}{(1-b)}
\frac{S_{L,M}}{B_{L,M}} \right]\;.
\label{EBS}
\end{equation}
Therefore
\begin{equation}
U_{L,M}\simeq
S^3_{L,M}B_{L,M} \frac{ (bz_c-1)^5}{z_c^2(1-b)}
\end{equation}
and (\ref{DeltaU}) along with (\ref{ZBS},\ref{XBS}) 
and $b=1-\beta$, yields
\begin{equation}
\Delta \simeq \frac{\beta(1-\beta)}{\rho-\beta}
\label{DBS}
\end{equation}
as in (\ref{res2}).

\noindent \underline{For $1/\alpha > z_c$ and $1/b > z_c$}\\
\noindent 
For this phase the evaluation of the integral (\ref{Uint})
is very similar to that outlined above for the previous phase, with
$\alpha$ replacing $b$. In carrying final integral over $w$
a factor of $-1$ is introduced due to the opposite directions
of the integral around $1/\alpha$ and $1/b$.
Therefore
one obtains the diffusion constant  by interchanging
$\beta$ and $1-\alpha$ in 
(\ref{DBS}) and multiplying by $-1$
\begin{equation}
\Delta \simeq \frac{ \alpha (1-\alpha)}{1-\rho-\alpha}\;.
\end{equation}
as in (\ref{res1}).

\noindent \underline{For $1/\alpha > z_c$ and $1/b < z_c$}\\
In this phase it turns out
that the dominant contributions to $U_{L,M}$ come from
one of $z,\tilde{z}$ at the pole $1/b$ and the other at the pole
$1/\alpha$ and $w$ at either of the two poles.
Carrying out the integrals in (\ref{Uint}) in sequence 
gives
\begin{eqnarray}
U_{L,M}&\simeq&\oint \frac{d w}{2\pi i}
\frac{w^{2L-2}}{(w-1)^{2M-1}} \frac{[1+E(0)(w-1)]^2}{(bw-1)^2(\alpha w-1)^2}
\oint \frac{dz}{2\pi i}
\frac{z^L}{(z-1)^M} \nonumber
\\
&&\times
\left[
 A_{L,M} \frac{(\alpha-1)(w -z)(\alpha w -1)}{\alpha (bz-1) z}
+
 B_{L,M} \frac{(b-1)(w -z)(b w -1)}{b (\alpha z-1) z}
\right] \nonumber
\\
&\simeq&
A_{L,M} B_{L,M} 
\frac{(b-\alpha)^2}{\alpha b}
\oint \frac{d w}{2\pi i}
\frac{w^{2L-2}}{(w-1)^{2M-1}} \frac{[1+E(0)(w-1)]^2 }{(bw-1)(\alpha w-1)}
\nonumber\\
&\simeq&
A_{L,M} B^3_{L,M} 
\frac{(b-\alpha)^3(1-b)}{\alpha }
\left[1+E(0)\left({1-b\over b}\right)\right]^2 
\nonumber \\
&&+
A^3_{L,M} B_{L,M} 
\frac{(b-\alpha)^3(1-\alpha)}{b }
\left[1+E(0)\left({1-\alpha\over \alpha}\right)\right]^2\;.
\label{UAB}
\end{eqnarray}
First consider $B_{L,M} \gg A_{L,M}$.
Then due to the form of $X_{L,M}$ (\ref{XAB}) in this phase
one has
\begin{equation}
E(0) \simeq - \frac{b}{1-b} \left[ 1 +\frac{(\alpha -b)}{(1-b)\alpha}
\frac{A_{L,M}}{B_{L,M}} \right].
\end{equation}
Both terms in (\ref{UAB}) contribute and one obtains
\begin{equation}
U_{L,M}
\simeq A^3_{L,M} B_{L,M} \frac{ (b-\alpha)^5}{\alpha^3 b (1-b)^2}
\left[ b(1-b) + \alpha(1-\alpha) \right]
\end{equation}
Then (\ref{XAB}), (\ref{ZAB}) and (\ref{DeltaU})
imply
\begin{equation}
\Delta \simeq \frac{\beta(1-\beta) +\alpha (1-\alpha)}{1-\beta -\alpha}
\label{DAB}
\end{equation}
as in (\ref{res3}).

\noindent In the case where $A_{L,M} \gg B_{L,M}$
\begin{equation}
E(0) \simeq - \frac{\alpha}{1-\alpha} 
\left[ 1 +\frac{(b -\alpha)}{b(1-\alpha)}
\frac{B_{L,M}}{A_{L,M}} \right]\;,
\end{equation}
however it turns out that
one obtains the same expression for the diffusion constant
(\ref{DAB}).

\end{document}